\documentclass[12pt]{article}%
\usepackage{hyperref}
\usepackage{cite}
\usepackage{amsmath}
\usepackage{amsfonts}
\usepackage{amssymb}
\usepackage{graphicx}
\usepackage{amscd}%
\ifx\pdfoutput\relax\let\pdfoutput=\undefined\fi
\newcount\msipdfoutput
\ifx\pdfoutput\undefined\else
\ifcase\pdfoutput\else
\ifx\paperwidth\undefined\else
\ifdim\paperheight=0pt\relax\else\pdfpageheight\paperheight\fi
\ifdim\paperwidth=0pt\relax\else\pdfpagewidth\paperwidth\fi
\fi\fi\fi
\begin{document}

\title{Inflaton vacuum fluctuations as dark matter and the potential V(phi) as dark energy}
\author{Max Chaves\thanks{Email: maximo.chaves@ucr.ac.cr}\\\textit{Escuela de Fisica}\\\textit{Universidad de Costa Rica}\\\textit{San Jose, Costa Rica}}
\date{August 18, 2018}
\maketitle

\begin{abstract}
It is shown, using quantum field theory in curved spacetime, how the expansion
of the universe during inflation produces an aggregate of particles and
inflaton vacuum fluctuations at a temperature of $5\times10^{17}$GeV and dense
enough to make reheating unnecessary. The standard calculation that predicts
the Hubble parameter has to be way smaller than the Planck energy is shown to
be fallacious: it applies the conservation of the perturbative curvature
$\mathcal{R}$ to a single inflaton fluctuation when it should be applied to
the energy density contrast of an aggregate. The quantum inflaton fluctuations
$\varphi$ are with respect to the classical value $\phi_{0}$ of the inflaton
field $\phi=\phi_{0}+\varphi$. Fluctuations $\varphi$ that have grown to the
size of the horizon, or a pair of virtual particles that are separated by a
distance the length of the horizon, are forced to become real and take energy
from the potential $V(\phi_{0})$. The slowing down of inflation is due to the
eventual domination of the continuously being created radiation over the
decreasing inflaton potential $V(\phi_{0}).$ It is not necessary at all for
the potential $V(\phi_{0})$ to go to zero. \emph{Since there is no need for
reheating the inflaton field }$\phi$ \emph{does not couple to matter (except
gravitationally).} After inflation, the fluctuations $\varphi$ quickly cool
down and can be described as dark matter. Now the inverse process begins to
occur. Inflaton fluctuations $\varphi$ that exited the horizon during
inflation begin reentering it after inflation's end. Then they are again
causally connected and have a probability of undergoing the inverse of the
quantum process they underwent before and give their energy back to the
potential $V(\phi_{0}).$ The $\varphi$ fluctuations are turning into
$V(\phi_{0}),$ which acts as dark energy and accelerates again the expansion
of the universe. The disintegration of a perturbation is a quantum jump of
cosmological size.


\end{abstract}

\section{Introduction.}

Understanding dark matter as a modification of Newtonian dynamics suffered a
blow due to the use of gravitational lensing.\cite{DM} There are related
covariant models that modify general relativity, sometimes with the addition
of a scalar field, so that gravity acts differently on large scales and mimics
dark matter. The recent observation of gravitational waves produced by the
binary neutron merger in the NGC 4993 galaxy,\cite{A1} simultaneously with the
observation of a short gamma-ray burst,\cite{Getal}, has made it possible to
conclude that the speeds of light and of gravitational waves are the same up
to one part in $10^{15}$.\cite{A2} As a result extraordinarily tight
constraints have been applied to the Horndeski and beyond-Horndeski theories
that were designed with dark matter (and sometimes dark energy) in
mind.\cite{LT,Betal,CV,SJ,EZ} An alternative explanation for dark matter are
particles. Much effort is being done in this area in laboratories and through
a variety of types of astronomical observations. So far the results have been
on the negative.\cite{Aetal,AGS,C,D,K} The conclusion would be, not that these
models have been disproved, since there is not enough evidence to reach that
conclusion, but that our limited knowledge certainly encourages fundamental
theoretical work.

Here we take a different approach to the problem of dark matter. We go back to
the inflationary epoch with the hope that it can shed light on the origin of
dark matter. As usual, we are going to assume that the inflationary epoch is
driven by the inflaton, a quantum scalar field $\phi$ with a potential energy
density $V(\phi)$. The pressure $p$ and density $\rho$ for this field in an
homogenous and isotropic universe are given by%
\begin{equation}
\rho=\dot{\phi}^{2}/2+V(\phi)\text{ and }p=\dot{\phi}^{2}/2-V(\phi).
\label{state}%
\end{equation}

The inflaton $\phi(t,\mathbf{x})$ is the sum of two terms: the classical field
$\phi_{0}(t\mathbf{),}$ which is a solution of the equations of motion
generated by the Lagrangian density $\mathcal{L}$ of the system, and the
quantum perturbative field $\varphi(t,\mathbf{x)}$:%
\begin{equation}
\phi(t,\mathbf{x})=\phi_{0}(t\mathbf{)+}\varphi(t,\mathbf{x).}
\label{inflaton}%
\end{equation}
Here $\phi_{0}(t)=\left\langle 0\left\vert \phi(t\mathbf{,x})\right\vert
0\right\rangle ,$ that is, $\phi_{0}$ is the vacuum expectation value of the
quantum field $\phi$, and has to satisfy the equation of motion of the
inflaton%
\begin{equation}
\ddot{\phi}_{0}+3H\dot{\phi}_{0}+V^{\prime}(\phi_{0})=0. \label{motion}%
\end{equation}
During the inflationary epoch the value of the Hubble horizon $H^{-1}$ remains
fairly constant except near the epoch's end. We assume a very small kinetic
energy term, so the inflaton acts as a perfect fluid with an equation of state
$\rho=-p.$ The two Friedmann equations that govern the inflationary expansion
(with no space curvature nor cosmological constant) are $-3\ddot{a}/a=4\pi
G(\rho+3p)$ and%
\begin{equation}
H^{2}=\left(  \dot{a}/a\right)  ^{2}=8\pi G\rho/3. \label{Friedmann}%
\end{equation}
One concludes that the solution is a fast-growing exponential $a(t)=\exp(tH),$
where $H=(8\pi G\rho/3)^{1/2},$ the Hubble parameter, is approximately
constant. It is assumed that $V(\phi_{0})$ has a small slope, so that the
value of $\phi_{0}$ is almost constant$.$ Notice that the potential
$V(\phi_{0})$ with the argument $\phi_{0},$ the classical part of $\phi,$ acts
as a repulsive cosmological constant.

In a Minkowski spacetime there are always vacuum fluctuations forming from the
quantum vacuum, but they soon disappear. But in a
Friedmann-Lema\^{\i}tre-Robertson-Walker (FLRW) spacetime the existence of a
Hubble horizon $H^{-1}$ results in the formation of a bath of Gaussian
fluctuations at Gibbons-Hawking temperature $T=H/2\pi.$\cite{B} Most of these
fluctuations are virtual, but the ones that are larger than the horizon do not
have enough time to disappear and thus become real. They grow in size and
acquire energy and become seeds for gravitational accretion of the dark matter
and particles that populate the universe. It is usually assumed that the
strength of an inflaton fluctuation determines the strength of an energy
density contrast $\delta\rho/\rho$ later on, after inflation.

We shall calculate the temperature and density of the particles and inflaton
vacuum fluctuations produced from the vacuum during inflation using standard
results of quantum field theory in curved spacetime.\cite{BD} It turns out
that during inflation there are no individual fluctuations to speak of;
instead, what is present is a thermal bath of fluctuations at a temperature of
about $5\times10^{17}$GeV, one such bath created every $e$-folding.
Furthermore, the bath is not simply made up of inflaton fluctuations, but of
all kinds of elementary particles. With all this matter there is no need to
have a reheating period at the end of inflation. These results rise another
question immediately: where are today all these inflaton fluctuations? The
thing is, they would be an excellent candidate for dark matter.\emph{ Since
reheating is not necessary anymore, we can assume the inflaton does not
interact with any other particle (except gravitationally), just like dark
matter does not.} Also, the amounts of dark matter and normal matter would be
comparable, as they are observed to be.

In Section 2 we will discuss in detail the topic of the relative size of the
modern universe density fluctuations and of the inflationary fluctuations.
This topic is closely related to the size of the Hubble parameter during
inflation, a point also discussed there. In Section 3 we calculate the
quantity of particles created from the vacuum during inflation using the
temperature of the thermal bath at the cosmological event horizon and its
spectral radiance. In Section 4 we study the transition between inflation and
the rest of the Big Bang, a period usually associated with reheating and preheating.

In Section 5 we give a summary of the paper and also discuss an interesting
offshoot of the idea that dark matter is made up of inflaton fluctuations.
Briefly, the idea there examined is as follows: Although both inflaton
fluctuations and elementary particles are created from the quantum vacuum,
their development in the FLRW universe is quite different. The size of
elementary particles is fixed, while the inflaton fluctuations grow
proportionally to the scale factor. If a perturbative inflaton's wavefunction
is larger than the horizon $H^{-1}$ it becomes impossible, due to causality,
for the perturbation to disappear back into the vacuum. Now, the fluctuation
has at least a size $H^{-1},$ maybe more (perhaps when created it was larger
than the horizon). It has to become real instantaneously, which implies that
the energy it needs has to be supplied to it locally. This energy must come
from the inflaton potential $V(\phi_{0}),$ which must then be locally very
slightly modified and weakened. The potential is imprinted with a negative of
the shape of the fluctuation. The fluctuation remains outside of the horizon
for some time and eventually, some time after the end of inflation, it goes
back inside due to the slowing down of the cosmic expansion and the increase
in size of the horizon. We shall argue in last section that, once the
fluctuation reenters the horizon, it can disintegrate (as causality does not
forbid it to do so anymore), and return to the background inflaton potential
$V(\phi_{0}).$ This process is a cosmological-size quantum transition, and is
equivalent to a bit of dark matter turning into gravitationally repulsive
material, or dark energy. With time more and more fluctuations disintegrate
and strengthen the background\ potential $V(\phi_{0}),$ until it again
dominates over inflaton fluctuations and matter particles, and the expansion
begins to accelerate.

\section{What the slow-roll and the perturbative curvature have to say about
the size of the Hubble horizon and the temperature of the universe during
inflation.}

In the previous section we introduced the idea that dark matter may be
composed of the same inflaton fluctuations that are believed nowadays to be
the source of the anisotropy observed in our universe. Usually this idea would
be rejected based on the consideration that inflaton quantum fluctuations
should have an intensity of the same order of magnitude as the energy density
contrasts $\delta\rho/\rho$ of the later universe, that is, of the order of
$10^{-5}$. This seems to be contradictory with the large amount of dark matter
present today in the universe, about 27\% of the energy density of the
universe. However, if we believe that during inflation large amounts of dark
and normal matter were created, then what can be concluded is that the small
fluctuations during inflation are simply energy density contrasts $\delta
\rho/\rho$ of an aggregate of inflaton fluctuations and normal matter, and the
fact that these may be small has no bearing on the amount of dark matter (and
other particles) back then, just in the same way that the $10^{-5}$ has no
bearing on the amount of dark matter today.

In this section we analyze in detail the implications of the smallness of the
fluctuations. A central point of our discussion is the size one can expect for
the Hubble horizon $H$ during inflation assuming the slow-roll regime. This is
relevant because in next section we are going to show, using the theory of
quantum fields in curved spacetime, that there is a particle and fluctuation
production from the vacuum during inflation. We shall call it "vacuum
production", for the sake of brevity. Since the temperature of the resulting
thermal bath is proportional to the horizon $H,$ we need to study what are the
possible values of $H$ in a slow-roll regime. To keep the exposition short, we
shall assume only one type of inflaton scalar field and only one potential,
the \emph{large field }or \emph{chaotic inflation }potential $V(\phi
_{0})=\frac{1}{2}m^{2}\phi_{0}^{2}$.

We now show that the slow-roll is perfectly compatible with a large value for
the Hubble parameter, $H\lesssim M_{P}$. Here $M_{P}$ is the Planck mass
$G^{-1/2}$. We shall assume:

\begin{enumerate}
\item The two fundamental inequalities of the slow-roll regime,%
\[
\epsilon=\frac{M_{P}^{2}}{16\pi}\left(  \frac{V^{\prime}}{V}\right)  ^{2}%
\ll1,\quad\eta=\frac{M_{Pl}^{2}}{8\pi}\left\vert \frac{V^{\prime\prime}}%
{V}\right\vert \ll1.
\]

\item That there are $N_{e}=60$ $e$-foldings between the horizon exit of the
earliest scales (the largest cosmological scales) and the end of the
inflationary regime.

\item That the potential density $V(\phi_{0})$ is $100$ times smaller than the
Planckian energy density $M_{P}^{4}$ (to avoid the quantum gravity regime).

\item The dynamical equations (\ref{state}), (\ref{motion}) and
(\ref{Friedmann}).
\end{enumerate}

These are all reasonable assumptions. For the chosen potential both
inequalities are equivalent to one,%
\[
\epsilon=\eta=M_{P}^{2}/4\pi\phi_{0}^{2}\ll1.
\]
Let us take $\epsilon=\eta=1/120,$ a small number chosen to obtain the
specific value $N_{e}=60$ $e$-foldings, as we shall see below. Then
\begin{equation}
\phi_{0}/M_{P}=\sqrt{120/4\pi}=3.1, \label{field}%
\end{equation}
and thus the classical field $\phi_{0}$ has to take super-Planckian values.
The relation between the quotient $\phi_{0}/M_{P}$ and $N_{e}$ is given by the
the following approximate calculation of the number of $e$-foldings underwent
by the universe from an initial time $t_{i}$ when the earliest scales (the
largest cosmological scales) exited the horizon to a final time $t_{f}$ at the
end of the inflationary regime :%
\[
N_{e}=\log\frac{a(t_{f})}{a(t_{i})}=\int_{t_{i}}^{t_{f}}Hdt=\frac{8\pi}%
{M_{P}^{2}}\int_{\phi_{f}}^{\phi_{i}}\frac{V}{V^{\prime}}d\phi\approx2\pi
\phi_{0}^{2}/M_{P}^{2}=60,
\]
where we have used (\ref{field}). In this model $\phi_{i}\gtrsim\phi
_{0}\gtrsim\phi_{f}$, and $\phi_{f}$ does not have to be zero or small. We
confirm the consistency of $N_{e}=60$ with $\epsilon=\eta=1/120$.

In order to ensure that the system does not enter the quantum gravity regime,
the condition $\frac{1}{2}m^{2}\phi_{0}^{2}\ll M_{P}^{4}$ must also be
imposed. Here $M_{P}^{4}$ is the Planck density. For the value of the inflaton
background field $\phi_{0}=3.1M_{P},$ this inequality leads to another
inequality, $4.8m^{2}/M_{P}^{2}\ll1.$ As a working hypothesis let us take
$4.8m^{2}/M_{P}^{2}$ to be one hundred times smaller than $1$, in which case
the value of $m$ comes out to be $m/M_{P}=1/22.$ The value of $H$ for this
value of $m/M_{P}$ can be found from Friedmann's equation (\ref{Friedmann})
for $H$:%
\begin{equation}
\frac{H}{M_{P}}=\sqrt{\frac{8\pi}{3}}\cdot\frac{1}{\sqrt{2}}\frac{m}{M_{P}%
}\frac{\phi_{0}}{M_{P}}=2.0\cdot\frac{1}{22}\cdot3.1=0.29 \label{Hubble}%
\end{equation}
This is a remarkable result because it puts the value of $H$ at $M_{P}$ or a
few orders of magnitude smaller. It shows that a value for the Hubble horizon
$H$ not much smaller than $M_{P}$ is fully compatible with the slow-roll. We
have not shown that $H$ has to be of this order, but that this possibility has
to be taken seriously. Let us call this the\textit{ high }$T$\textit{ choice
of parameters. }

We shall see in the next section that for a Hubble horizon $H$ not too much
smaller than the Planck mass $M_{P},$ the vacuum production is enough to
account for all the matter in the universe.

Soon after inflation was introduced it was noticed that there was a
fundamental quantity, the curvature perturbation $\mathcal{R}$ in the comoving
reference frame, which has a constant value from the time it exits the horizon
during the inflationary epoch, until the time it reenters it in the the modern
universe.\cite{B-L,W} This quantity $\mathcal{R}$ allows us to relate the size
of fluctuations during the inflationary period with the size of fluctuations
in the modern universe, which is is $\Delta_{\mathcal{R}}=5.0\times10^{-5}$
(quoting significant figures common to those reported by the different
groups).\cite{KB}

It is usually assumed that the origin of the inhomogeneities of the universe
are the quantum fluctuations of the inflaton field. It is assumed that during
inflation there is no matter in the universe other than the inflaton quantum
field, and that the subsequent structure of the universe is determined by the
two-point quantum correlation $\left\langle 0\left\vert \varphi(t,\mathbf{x)}%
^{2}\right\vert 0\right\rangle $ of this field evaluated at the same event.
The ensemble for this correlation is the usual quantum one. But the
possibility of using the quantum correlation as a source for the
inhomogeneities of the matter distribution in the universe is predicated on
the absence of matter in the universe during inflation, other than the quantum
inflaton. If there is already an aggregate of particles in the inflationary
universe the correct way to calculate the quantum correlation would be using
finite-temperature quantum field theory and the density matrix, and a mixed
quantum and thermal ensemble.

Let us make the usual assumptions that during inflation the universe is empty
or at such a low temperature that we can ignore whatever particles may exist
with the exception of the quantum inflaton $\varphi$. To achieve these
conditions within the slow-roll it is necessary to assume that the energy
density $V(\phi_{0})$ of the inflaton is far smaller than the Planck energy
density $M_{P}^{4},$ so as to obtain a smaller value for $H$ and a low
temperature $T.$ For the potential $V=\frac{1}{2}m^{2}\phi_{0}^{2}$ and the
commonly used values $N_{e}\sim60$ and $\phi_{0}\sim3M_{P}$, we shall see
below that it is necessary to assume that $V/M_{P}^{4}\sim10^{-11}$ in order
to obtain a value for $\mathcal{R}$ consistent with astronomical measurements.
Let us call this the $T=0$\textit{ choice of parameters }and sketch the
mathematical derivation of the size of $H$. Outside the horizon the
perturbative curvature $\mathcal{R}$ is related\cite{LL} to the inflaton
quantum field $\varphi$ by%
\begin{equation}
\mathcal{R}\approx-\frac{H}{\dot{\phi}_{c}}\varphi. \label{link}%
\end{equation}
The quantities $H$ and $\dot{\phi}_{c}$ are fairly constant classical numbers.
(This calculation can be done more accurately,\cite{Wc} but it is not
necessary for our purposes to do so.)

To find the effect of the inflaton quantum field $\varphi$ on the energy
distribution of the universe taking $T=0$ we must calculate the two-point
correlation for a single location $\left\langle 0\left\vert \varphi
(\mathbf{x)}^{2}\right\vert 0\right\rangle $, under the assumption that the
field has left the horizon. The result is: \cite{LL}%
\begin{equation}
\left\langle 0\left\vert \varphi(\mathbf{x)}^{2}\right\vert 0\right\rangle
=\left(  \frac{H}{2\pi}\right)  ^{2}, \label{correlation}%
\end{equation}
where the $H$ is to be evaluated at the time of horizon's exit. The
perturbative curvature $\mathcal{R}$ is also a quantum field, but, as
(\ref{link}) shows, it is the same quantum degree of freedom as $\varphi.$
With the help of (\ref{link}) we conclude that%
\[
\left\langle 0\left\vert \mathcal{R}(\mathbf{x)}^{2}\right\vert 0\right\rangle
=\left\langle 0\left\vert \varphi(\mathbf{x)}^{2}\right\vert 0\right\rangle
\left(  \frac{H}{\dot{\phi}_{0}}\right)  ^{2}=\left(  \frac{H^{2}}{2\pi
\dot{\phi}_{0}}\right)  ^{2}.
\]
For this $T=0$ case we define the amplitude of $\mathcal{R}$ to be%
\begin{equation}
\Delta_{\mathcal{R}}\equiv\left(  \left\langle 0\left\vert \mathcal{R}%
(\mathbf{x)}^{2}\right\vert 0\right\rangle \right)  ^{1/2}=\frac{H^{2}}%
{2\pi|\dot{\phi}_{0}|}. \label{R}%
\end{equation}
From this equation it is possible to find the value of the mass $m.$ In order
to achieve this we substitute the value $\Delta_{\mathcal{R}}=5.0\times
10^{-5}$ in (\ref{R}), then also substitute in that same equation the value of
$H$ using the Friedmann equation (\ref{Friedmann}) and the value of $\dot
{\phi}_{0}$ using the inflaton's equation of motion (\ref{motion}) (neglecting
the $\ddot{\phi}_{0}$ term). This way we find $m/M_{P}=1.3\times10^{-6}.$
Plugging this result into the Friedmann equation we immediately get%
\begin{equation}
\frac{H}{M_{P}}=\sqrt{\frac{8\pi}{3\cdot2}}\frac{m}{M_{P}}\cdot\frac{\phi_{0}%
}{M_{P}}=8.2\times10^{-6}. \label{H}%
\end{equation}
This resulting value of $H$ is a lot smaller than the one of the \textit{high
}$T$\textit{ choice}. Let us verify that our estimation of $V(\phi_{0}%
)/M_{P}^{4}\sim10^{-11}$ was correct:%
\[
\frac{1}{2}m^{2}\phi_{0}^{2}/M_{P}^{4}=\frac{1}{2}\frac{m^{2}}{M_{P}^{2}}%
\frac{\phi_{0}^{2}}{M_{P}^{2}}\sim10^{-11}.
\]

We have seen that in order to avoid the quantum gravity regime the condition
$\frac{1}{2}m^{2}\phi_{0}^{2}\ll M_{P}^{4}$ must be satisfied, but that there
is leeway in how much smaller one can make one quantity than the other. The
calculations for the two choices presented here, one taking the ratio of the
two densities to be $1/100$ and the other to be $1/10^{11},$ involve
completely different physics. It is fallacious to discard the \textit{high
}$T$\textit{ choice} on the basis of a calculation that is based on physics
that belong to the $T=0$\textit{ choice}, but that is precisely what is done
when equation (\ref{R}) is taken to always imply (\ref{H}), that is, a very
small Hubble parameter. Equation (\ref{R}) was derived on the assumptions of a
very low temperature and $V/M_{P}^{4}\sim10^{-11}.$ As we have seen, (\ref{R})
comes directly from (\ref{correlation}), an equation which is basically
irrelevant in the high $T$ choice. For this high $T$ choice all that the
smallness of $\mathcal{R}$ implies is that the fluctuations of the inflatons
and particles aggreggate are small, too, but it gives no direct information
about $\dot{\phi}_{0}$ or $H.$ For the high $T$ choice there is no equivalent
to equation (\ref{R}).

We have seen that taking natural values for the slow-roll can result on a
large value for $H.$ This in turn determines a large Gibbons-Hawking
temperature and a copious production of particles, a phenomenon that gives an
explanation for the origin of the matter of the universe. As more and more
particles are produced the ratio of the radiation density $\rho$ to the
potential $V(\phi_{0})$ increases and eventually inflation ends. This is a
natural mechanism to cause the end of inflation.

\section{Calculation of the quantity of particle production from the vacuum
during inflation.}

In the quantum vacuum of Minkowski spacetime, particles are constantly
appearing and disappearing, but it is impossible for them to become real since
the principle of conservation of energy forbids it. But in a spacetime that
possesses a causal horizon, such as the FLRW, if a pair of virtual particles
become separated by a distance larger than the horizon (the Hubble horizon in
this case), they will not have time to reunite and are therefore forced to
become real particles. Similarly, if a fluctuation of the inflaton field
$\varphi$ becomes equal or larger than the Hubble horizon, the causal
microprocesses necessary to take the fluctuation back into nothingness do not
have enough time to act and the fluctuation necessarily has to remain in
existence. The energy $\Delta E$ available for vacuum production is given by
the uncertainty principle $\Delta E\Delta t\approx1,$ and one gets the
approximate result $\Delta E\approx\Delta t^{-1}\approx H.$

It is to be expected that this energy $\Delta E$ has to come from the
potential energy $V(\phi_{0}),$ so that the value of $\phi_{0}$ has to change
by a small amount $\delta\phi_{0}=\phi_{0}^{\prime}-\phi_{0},$ where $\phi
_{0}^{\prime}$ differs from $\phi_{0}$ only locally. An energy $\Delta E$ has
become available and equal the integral over space of $V(\phi_{0})-V(\phi
_{0}^{\prime}).$ We assume that during inflation this happens constantly and
ubiquitously so that the background field remains basically homogeneous, of
the form $\phi_{0}(t).$

There is a large literature on thermal radiation baths present in accelerated
frames and gravitational fields.\cite{BD} Under some circumstances these
particles should become real, the best well-known example being that of the
radiation emitted by a black hole. It was observed in \cite{GH} that a de
Sitter spacetime with a repulsive cosmological constant $\Lambda$ contains a
cosmological event horizon with a particle thermal bath. Gibbons and Hawking
succeeded in finding the temperature of the particle bath in terms of the
surface gravity $\kappa$ of the cosmological event horizon as seen by an
observer stationed there. Their result was%
\begin{equation}
T=\kappa/2\pi=\sqrt{\Lambda/3}/2\pi=H/2\pi. \label{G&H}%
\end{equation}
If instead one assumes that inflation is caused by the inflaton field, then,
according to a Friedmann equation (\ref{Friedmann}), the Hubble horizon would
be given by $H^{2}=8\pi G\rho/3.$ One of the branches of a de Sitter spacetime
is equivalent to a FLRW spacetime with an increasing exponential scale factor.
Since both expansions are physically equivalent we conclude that the
temperature at the cosmological event horizon must given by $T=\sqrt{8\pi
G\rho/3}/2\pi.$

Physical consequences of the thermodynamics of cosmological event horizons
(and other types of horizons, too) have been studied in \cite{P}. One very
natural idea put forward there is that the energy of the radiation produced
from the vacuum must come from the source of the gravitational fields or
accelerations involved in the creation of the horizons. If we assume that the
accelerating expansion of the universe is due to the inflaton field, then the
energy of this field must be weakened by vacuum production. This situation was
studied in detail in \cite{CB} and the dynamic development of an accelerated
expanding universe was described. Here we are not going to concern ourselves
with the time dependence of the Hubble horizon. We are going to assume the
slow-roll regime and take the horizon $H$ (and thus the temperature) to be constant.

We want to know how much radiation is being created per unit volume per unit
time during inflation. We assume that there is a thermal bath at the
cosmological event horizon, at a temperature $T$ given by formula (\ref{G&H})
above. We take the event horizon to have a spherical shape with diameter
$H^{-1}.$ In the surface of the sphere we take a small area $dA$ \emph{and
calculate the energy flux leaving the sphere through that area} (for a certain
type of particle) using the spectral radiance:%
\[
B=\frac{2h\nu^{3}}{c^{2}}\frac{1}{\exp(h\nu/kT-1)}=\frac{\omega^{3}}{2\pi^{2}%
}\frac{1}{e^{\omega/T}-1},
\]
where the last expression on the right is in natural units, $\hbar=c=k=1.$ The
spectral power flux $P_{\omega}$ passing through the small area $dA$ is%
\[
P_{\omega}\,dA=dA\int_{0}^{\pi/2}\sin\theta\cos\theta\,d\theta\int_{0}^{2\pi
}d\phi\,B=\pi B\,dA,
\]
a calculation done using "Lambert's cosine law", since the flux leaving the
sphere through $dA$ has a $2\pi$sr spread. To find the total energy density
flowing out of the sphere we have to integrate over the surface of the sphere
(which is done simply by multiplying by its area $A=4\pi(H^{-1}/2)^{2}$)$,$
and over all possible frequencies using the differential $d\nu=d\omega/2\pi$:%
\begin{equation}
\Phi=A\int_{0}^{\infty}P_{\omega}\,d\omega/2\pi. \label{integral}%
\end{equation}

In Refs. \cite{GH,P,CB} it is assumed that the Hubble horizon of an
exponentially accelerated expansion is a true cosmological event horizon and
that \emph{all} the particles in the thermal bath do become real, in which
case the upper limit of the definite integral (\ref{integral}) should be
$\infty.$ However, there does not seem to be a mandatory reason for the
virtual particles with wavevectors $k/a>H,$ which have not exited the horizon,
to become real. They can, instead, go back to the vacuum within the period
allowed by the uncertainty principle, so that the upper limit in
(\ref{integral}) should be $H.$ In any case, taking infinity instead of $H$ as
the upper limit of the integral only increases its value by 13\%. We will use
infinity as the upper limit simply because it results in an exact
Bose-Einstein integral. To perform the integration in (\ref{integral}) we
proceed as follows:%
\[
\Phi=\frac{A}{4\pi^{2}}\int_{0}^{\infty}\frac{\omega^{3}d\omega}{e^{\omega
/T}-1}=\frac{AT^{4}}{4\pi^{2}}\int_{0}^{\infty}\frac{x^{3}dx}{e^{x}-1}%
=\frac{\pi T^{2}}{240}.
\]
This is the flux of energy flowing out of the sphere$.$

By the symmetry of the physical problem the flux of energy leaving the sphere
has to equal the flux of energy entering it. The power density inside is the
amount of energy entering per unit time, divided by the volume $V=\frac{4}%
{3}\pi(H^{-1}/2)^{3},$ or:%
\[
\Phi/V=\pi T^{2}/240V=\pi^{2}T^{4}H/10.
\]
Furthermore, the energy density $\rho_{e\text{-folding}}$ created in one
$e$-folding would be the power inside the sphere times $H^{-1},$ leading to
the result%
\begin{equation}
\rho_{e\text{-folding}}=\pi^{2}T^{4}/10. \label{density}%
\end{equation}
We mention for purposes of comparison this density is slightly larger than the
energy of a photon gas at temperature $T$, which is $u_{\gamma}=\pi^{2}%
T^{4}/15.$ Notice the production of radiation due to the expansion of the
universe is a dissipative mechanism.

Finally, let us assume that there are about 120 degrees of freedom in the high
energy standard model. Every fundamental particle must have its own thermal
bath (unless the mass of a particle is larger than the temperature), since the
arguments for the existence of a bath for a type of particle are completely
generic, given a fundamental particle. The density (\ref{density}) is for
photons, which have two helicities. For the standard model we should then have
a density 60 times larger. Thus, a density%
\[
\rho=6\pi^{2}T^{4}%
\]
is being created every $e$-folding. The temperature can be calculated from $H$
as given by (\ref{Hubble}) and (\ref{G&H}) and is%
\[
T=0.3M_{P}/2\pi=5\times10^{17}\text{GeV.}%
\]

This temperature is higher than that needed for grand unification symmetry
breaking, even assuming supersymmetry.

\section{The transition between inflation and the rest of the Big Bang.}

It is usually assumed that the slow-roll lasts about 60 $e$-foldings, and that
then (or soon after) reheating begins.\cite{AFW,ASTW} It is assumed that
during reheating there is a total conversion of the potential energy of the
inflaton into particles so that at the end of inflation $V=0.$%
\cite{KLS,Fetal,AHKK} The purpose of the reheating phase is to explain the
origin of the matter of the universe. Since reheating requires a strong
interaction of the inflaton field with other particles, it seems necessary the
inflaton potential should be zero by the end of the reheating period, as
otherwise it would interact with the particles in the universe later on, in
processes that have not been observed. But if one assumes that vacuum
production results in large quantities of inflaton fluctuations and particles
being created throughout the slow-roll, then there is no need to assume a
reheating period at all. Matter is created beforehand from the vacuum by
quantum gravity effects.

During inflation the inflaton rolls slowly down the potential $V(\phi_{0}),$
spending the energy its is gaining in sustaining the production from the
vacuum. Towards the end of the slow-roll the domination of the inflaton is put
into question by the accumulated particles and inflaton fluctuations that have
been produced from the vacuum, and by the fact that the potential $V(\phi
_{0})$ itself has diminished. During this transition period the Hubble horizon
$H^{-1}$ begins to increase but there is still vacuum production (colder now
since the $H$ is smaller). Eventually radiation dominates, but there would
still be potential $V(\phi_{0})$ left, in a quantity comparable to the amount
of inflaton perturbations $\varphi$ and particle radiation. Since it does not
interact with matter it would be invisible today (except gravitationally).

After the end of inflation there would be a large quantity of inflaton
fluctuations, comparable with the quantity of particle radiation present at
that same time. These fluctuations do not interact with matter at all (except
gravitationally), and their kinetic energy term has a $1/a^{2}(t)$ factor so
they rapidly cool down with the expansion of the universe. They are good
candidates for dark matter.

\section{Summary and a possible role of the inflaton potential $V(\phi_{0})$
as dark energy.}

We have shown that during inflation there is a production from the quantum
vacuum of an energy density of $6\pi^{2}T^{4}$ per $e$-folding, due to quantum
gravity effects. The temperature is high, of the order of $5\times10^{17}$GeV,
enough to break the grand unification symmetry, even assuming supersymmetry.
The calculation that restrained the value of the Hubble parameter $H$ to be
low was shown to be invalid, since it is based on an incorrect application of
the conservation of the perturbative curvature $\mathcal{R}$. The fallacy is
to apply, during inflation, this conservation law to a single inflaton
perturbation. Since what is present then is, already, an aggregate, the
conservation law has to be applied to an energy density contrast $\delta
\rho/\rho.$

The large quantity of matter already produced makes reheating, and thus the
coupling of the inflaton to matter, unnecessary. In our picture inflation ends
when the fluctuations and particle radiation dominate over the potential
$V(\phi_{0}),$ and the universe enters a period of radiation domination. At
the end of inflation there will be a hot aggregate of particles and inflaton
fluctuations, and some potential $V(\phi_{0})$ left. The process does not
increase very much the value of $\dot{\phi}_{0}^{2}$ so that the inflaton
background field $\phi_{0}$ satisfies an equation of state $p\approx-\rho$. It
is our contention in this paper that the inflaton vacuum fluctuations, which
do not interact with matter except gravitationally, are the dark matter
observed in the universe.

Fluctuations with scales $k$ were created during the $e$-foldings of the
slow-roll, and each one came out of the horizon when $k/a=H,$ during the
inflationary epoch. After inflation's end the fluctuations have been
reentering the horizon, one by one, with the scales of smaller physical size
reentering first, larger ones last. When a virtual vacuum fluctuation, during
inflation, reaches the Hubble horizon, it has to become real. It, along with
the metric field $g_{\mu\nu}$ which is part of the solution, has to transform
locally (within small distances that are still causally connected) in order
for it to become a classical solution of the equations of motion. To be able
to do this, it must locally take energy from $\phi_{0},$ and in so doing leave
a small dent in $V(\phi_{0})$. The fluctuation leaves a negative image of
itself in the potential $V(\phi_{0}).$ Time passes and the inflaton
fluctuation eventually reenters the horizon. Once this happens, it is
possible, since causality is no longer an issue, for the fluctuation to
undergo the inverse of the quantum process that originally created it, and go
back to the vacuum. The dented volume, the fluctuation's negative image that
it left in $V(\phi_{0}),$ has expanded at the same rate as the fluctuation and
they are sharing the same location. It is as if there were a puzzle, and one
piece of it is lifted up; then if both the piece and the puzzle expand
together at the same rate they should still fit. There would be a quantum
amplitude for the fluctuation to go back to the vacuum. \emph{Since the
quantum process occurred in one direction in time, there should be a finite
probability for it to occur in the opposite direction.}

The fluctuation that has reentered the horizon is not obliged to go back to
the vacuum;\ it only has a probability of doing so. This inverse process is a
disintegration and, as such, it has a half-life. The process for a fluctuation
to go back into the vacuum can take a long time because of two completely
different reasons:

\begin{itemize}
\item It is possible that either the fluctuation $\varphi$ or the potential
$V(\phi_{0})$ have being distorted gravitationally by other objects before
reentry, in which case the quantum amplitude would become smaller or zero,
since the path integral is strongly inhibited by the resulting gradients.

\item Even if the quantum process of vacuum reabsorption of the inflaton
fluctuation actually begins to take place, the time scale of the quantum
transition is large because of the cosmological distances involved. Depending
on the scale involved, the quantum process of disintegration could take
hundreds or thousands of millions of years.\noindent
\end{itemize}

As more and more fluctuations enter the horizon and become eligible for
disintegration back into $V(\phi_{0})$, the chance for some of them to go back
to being part of the potential $V(\phi_{0})$ increases, and eventually many
will. This potential $V(\phi_{0})$ satisfies $p\approx-\rho,$ precisely as has
been observed nowadays for dark energy.\cite{AGS,CKW} As a result of the
disintegrations, the amount of potential $V(\phi_{0})$ will increase and
eventually dominate over the $\varphi$ fluctuations, and the expansion of the
universe begins accelerating again.

The inflaton has dominated the evolution of the universe. Initially the
potential $V(\phi_{0})$ was the direct cause of inflation. Then, in the form
of vacuum fluctuations, it is dark matter and helped the formation of
structure. Later, the potential $V(\phi)$ grew again and became dark energy.

{\Large \noindent}

\end{document}